# Observations of zero electrical resistance of Au-Ag thin films near room temperature

M. K. Hooda[1], P. Kumar[2], Viswanath Balakrishnan[2], and C. S. Yadav[1*]

[1]School of Basic Sciences, Indian Institute of Technology Mandi, Mandi-175005 (H.P.) India
[2]School of Engineering, Indian Institute of Technology Mandi, Mandi-175005 (H.P.) India
Email: shekhar@iitmandi.ac.in

Recent observations of superconducting like transition at 286 K in Ag and Ag nanostructures by Thapa *et al.* (arxiv: 1807.08572) have rekindled the hope for room temperature superconductivity under ambient conditions. We also investigated the electrical properties of Ag-Au nanostructure in the form of thin film grown on $SiO_2$/Si substrate by DC sputtering and observed signature of zero resistance in the temperatures range of 243 K to 275 K. While the observed electrical resistance of samples shows intriguing and perplexing behavior under temperature cycling and external magnetic field; the large spatial inhomogeneity present in thin film hampers the reproducibility indicating the stability issues associated with superconducting like phase.

The superconductivity (SC) is a macroscopic quantum phenomenon where material shows zero electrical resistance and perfect diamagnetism when cooled below a certain critical temperature [1]. The SC has fascinated the scientific community since its discovery in 1911 by K. Onnes due to its exotic physics and technological applications [1-6]. However certain critical parameters of SC such as temperature ($T_C$), magnetic field ($H_C$) and critical current ($J_C$) put the limit on technological usage of superconductors [3]. In the last four decades, several new superconductors have been discovered and SC transition temperature ($T_C$) has jumped up to few hundred Kelvin [3-10]. The Bardeen-Cooper-Schrieffer limit of 23 K was also broken by La-Ba-Cu-O oxide superconductor discovered by J. Bednorz and K. Mueller in 1986 [7]. This discovery further pushed $T_C$ of ~ 134 K in Hg- Ba-La-Cu-O, which is maximum $T_C$ achieved at ambient pressure till now [8]. In 2015, A. P. Drozdov *et al.* reported record $T_C$ of 203 K in the compressed $H_2S$ at the pressure of 155 GPa [10]. However room temperature SC has been elusive until now. During the last few decades nanostructure engineering has been playing a key role to tune the SC parameters as nanostructures can drastically modify the SC characteristic lengths (coherence length and penetration depth) due to quantum confinement and can also possibly improve $T_C$, critical current and magnetic field etc. [11].

Back in July 2018, claim for room temperature SC at ambient pressure in nanostructures (NS) comprising the silver (Ag) particles embedded in the gold (Au) matrix was reported on arxiv (arxiv:1807.08572v1) by D. K. Thapa and Anshu Pandey of Indian Institute of Science (IISc) Bangalore, India [12]. Considering the potential applications of room temperature (RT) SC, their results created sensation and excitement among the scientific community. Following the above claim and excitement for the verification of RT SC in NS, A. Biswas *et al.* prepared the thin films of Ag-Au NS using pulsed laser deposition technique on silicon and quartz substrates and reported electrical resistivity and magnetization measurements on NS[13]. The electrical resistivity (ρ) showed metallic behavior and in magnetization data, no signature of Meissner effect (perfect diamagnetism) was observed completely ruling out the possibility of RT SC [13]. Despite the negative evidence of RT SC [13] and consistent repeated pattern of noise in Au-Ag NS, which was highlighted by MIT physicist Brian Skinner [14], the debate and interest regarding the possibility of RT SC in Au Ag NS is still quite intense. Very recently lanthanum hydride ($LaH_{10}$) was reported to be superconducting at 250 K under very high pressure of 170 GPa[15]. The requirement of such a high pressure and difficulty in synthesis of hydride samples, pose challenges in the practical use of SC for potential applications. The group from IISc Bangalore D. K. Thapa *et al.* has recently updated the details of sample synthesis with more convincing experimental results in support of their previous observation (modified version arxiv:1807.08572v3). These results on Au-Ag NS by D. K. Thapa *et al.* has revived the hope of RT SC at ambient pressure [16]. Therefore it is quintessential to look for materials that can be easily synthesized and can show SC at ambient conditions. Inspired by these reports, we investigated the electrical resistance of thin films of Au-Ag thin films grown on different substrates ($SiO_2$/Si, glass and quartz) by DC sputtering.

## Material and Methods

$SiO_2$/Si substrate ($1 \times 1$ $cm^2$) was used to deposit the thin film of Au and Ag. Before the deposition, $SiO_2$/Si substrate was cleaned neatly using acetone and iso-propyl alcohol followed by 1 hr treatment in piranha solution (1:3 ratio of $H_2O_2$ and $H_2SO_4$). DC magnetron sputtering is used to make thin film of Au and Ag in sequential manner to produce embedded nanostructure of Ag in Au matrix. Initially, Au has been DC sputtered at rate 0.5 Å/sec with the base pressure of $5 \times 10^{-6}$ mbar pressure at room temperature. After completing the Au deposition of 50 nm thick, we increased the substrate temperature from room temperature to high temperature in the range of 300 °C to 500 °C while keeping environment intact. After reaching the specific substrate temperature, the already deposited Au thin film on $SiO_2$/Si was kept in hold at the same temperature for 30 min. Then Ag thin film of 5 nm thickness has been deposited using DC sputtering. Microstructural characterizations of as grown thin films were carried out using field emission scanning electron microscope (FESEM- FEI-Nova NanoSEM 450). Electrical measurements were carried out on both continuous and patterned thin films. Patterned thin film has been deposited using hard mask with dimension of 5 mm × 0.5mm.

DC Sputtered thin film of Au-Ag has been pre-analysed for their electrical characteristics using the probe station before heading to Quantum Design make PPMS measurements. Keysight-B2902A SMU (Source and Measure Unit) has been used to analyse the resistance variation of Au-Ag thin film over $SiO_2$/Si substrate of $1 \times 1$ $cm^2$. Two probes of tungsten have been used for performing I-V measurements by systematically varying the separation between W probes. For electrical resistance measurement in PPMS, we used linear four probe geometry and electrical contacts were made using silver paste.

## Results and Discussion

Au-Ag thin films on $SiO_2$/Si and other substrates were grown by DC sputtering at different temperatures as described in experimental methods. While the Au film is deposited at room temperature, we carried out Ag deposition at high temperature to enable microstructural accommodation of Ag particles in Au matrix. We limited the maximum temperature for Ag deposition on Au film to 500 °C in order to avoid dewetting of Au-Ag thin films at high temperature. Two different temperatures, 300 °C and 400 °C worked perfectly and resulted in the formation of good thin film containing Au-Ag nanostructures. SEM observation confirms the formation of polycrystalline microstructure containing bimodal grain size distribution of Au-Ag nanostructures (Figure 1a & 1b). This suggests that the overall Au-Ag thin films contain both fine grained (20 – 50 nm) and coarse grained (200 – 350 nm) microstructures. We measured the electrical resistance on several of Au and Ag nanostructured thin films prepared on different substrates.

We have shown the temperature dependence of Au-Ag film grown on silicon substrate (S1) in Fig. 2, 3. As shown in the Fig. 2, electrical resistance (measured longitudinal current I = 5 mA) is quite noisy above room temperature before showing consistent drop from 290 K to zero value at T = 275 K. However the large variation in the transition temperature values (240- 275 K) in various cooling and heating cycles (Fig 3a, 3b) raises the doubt on the possibility of superconducting nature of this transition. The range and variation in $T_C$ values for Au-Ag thin film is similar to the fine Ag nanoparticles embedded in Au matrix synthesized by wet chemical methods reported by Thapa *et al.* for different samples [12, 16]. Generally, SC materials exhibits detrimental effect of magnetic field on $T_C$, and Thapa *et al.* has observed the same up to H = 5 T field in their resistance and magnetic susceptibility measurements [16]. Nevertheless, the transition in our film does not obey this trend. The resistance measured at H = 5, 14 T fields neither show the loss nor the consistent decrease in transition temperature (Fig 3a, 3b). The erratic variation in transition temperature upon cooling and heating cycles in the presence or absence of magnetic field raises the questions on the stability of Au-Ag thin films over multiple heating-cooling cycles and also on the nature of plausible SC in these compounds. However considering the observed robustness of the transition even at 14 T magnetic field, these materials promise very high technological advantages.

It is to mention here that the transition width on our film is quite large (20 - 40 K), and resistance decreases to the lowest detectable limit of the Quantum Design PPMS. It is possible that the distribution of Ag within Au matrix, coupling of Ag-Au NS and the inter-grain connectivity are very sensitive to the temperature and follow a random behavior during cooling and heating cycles. This random dynamics destabilizes the plausible zero resistance state and leads to the variation in transition temperature. We tried to measure the low field dc susceptibility on the films, but we did not observe any diamagnetic signal around the zero resistance transition, and data remains very noisy. It is possible that the anticipated diamagnetic signal originating from any transient superconducting phase (if it is present) remains very small, as the total mass of the Au-Ag NS film used for the measurement is in the order of micrograms instead of 60 to 90 mg of samples used by Thapa *et al*. [16].

We measured electrical resistance on the other Au-Ag NS films grown on glass and quartz substrates and pure Au film grown on quartz substrate, under similar temperature conditions. To our surprise, we did not observe zero resistance in any other film (Fig. 4), and our results are similar (metallic only) to that reported by pulsed laser deposited Au-Ag films by Biswas *et al.* [13]. It is to mention that we observe metallic behavior with some random jumps in resistance (See Supplementary figure S1) for some films.

To further check the inter-grain connectivity of the film, we measured current-voltage (I-V) characteristics on a film (without any transition at RT) using two probe method. The probes were placed at different spacing to check the resistance on different parts of the films. We observed large scatter in the resistance values measured at different points. Surprisingly the observed resistance variation did not show the expected trend with dimensions (separation distance). The observed variation in the resistance values suggests the special inhomogeneity in composition and/or microstructure of Au-Ag thin film. This variation could be realised based on the microstructural/compositional changes with respect to the accommodation of Ag nanoparticles in Au matrix. Scatter in the resistance values are evident from the I-V curves measured for various regimes and different samples as shown in Fig 5. The observed Ohmic behaviour confirms the highly conducting path formation over the Au-Ag film. Nevertheless I-V plots obtained from different regions indicates high resistive path as well. The observed scatter in the electrical transport and its possible origin of spatial heterogeneity present in the Au-Ag thin film is one of the main reasons for the difficulty in reproducing the zero resistance behaviour. It is possible that the embedded Ag in Au matrix with specific composition stabilizes locally and contributes for forming superconducting like channels as reported recently [16].

**Conclusions**

We observed zero electrical resistance near room temperature (T = 240 - 275 K) in one of the Au-Ag nanostructured thin films deposited on $SiO_2$/Si substrate. The variation in transition temperature upon cooling and heating cycles and inconsistence dependence of transition temperature on magnetic field on our measurements do not support the clean superconducting behavior. This might be due to the instability issues associated with Au-Ag thin films during multiple heating-cooling cycles. The observed transition to zero resistance, albeit shown in one of the films only, is quite remarkable and requires more investigations on similar system. The present investigation along with previous report by Thapa *et al*. suggests the presence of an exotic and hidden superconducting like phase in Au-Ag nanostructures. The observed issues of instability and reproducibility may be associated with structure, lattice dynamics and/or spatial and compositional variations of Ag and Au nanoparticles in these systems. Further studies are required to better control the samples quality and understand the nature and origin of SC in these systems to harness the long awaited technological applications.

**Acknowledgement:** The authors acknowledge Advanced Material Research Center (AMRC), IIT Mandi for the experimental facilities. The financial support from the IIT Mandi and DST-SERB project YSS/2015/000814 is also acknowledged.

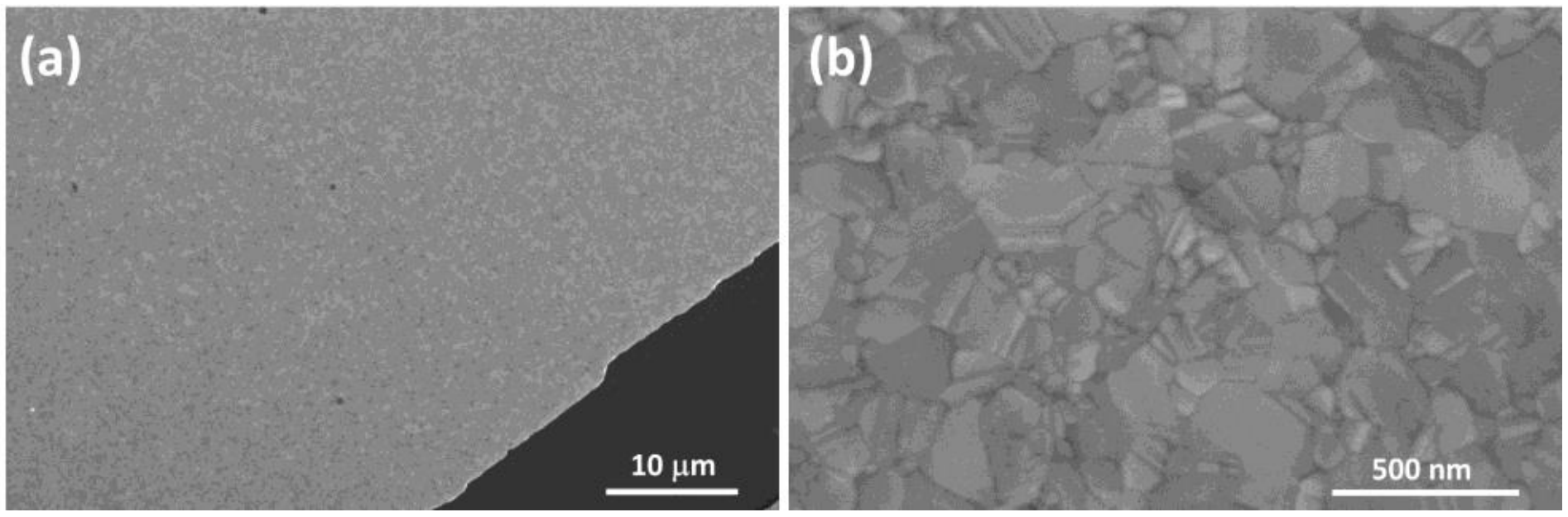

Fig 1: Low (a) and high (b) magnification SEM images of as grown Au-Ag thin film on $SiO_2$/Si substrate. Polycrystalline microstructure of Au-Ag thin film containing bimodal grain size distribution is evident from high magnification image.

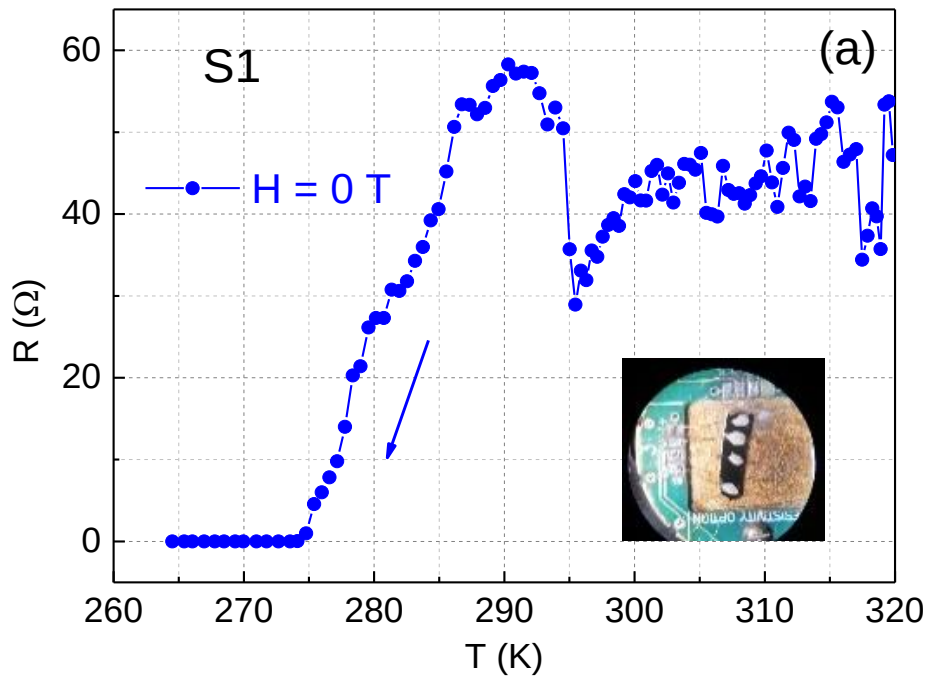

Fig. 2: R versus T for Ag-Au nanostructured film on $SiO_2$/Si substrate showing transition to zero resistance value 275 K under zero magnetic field. Image of the measured film mounted on the electrical resistance puck is also shown in figure.

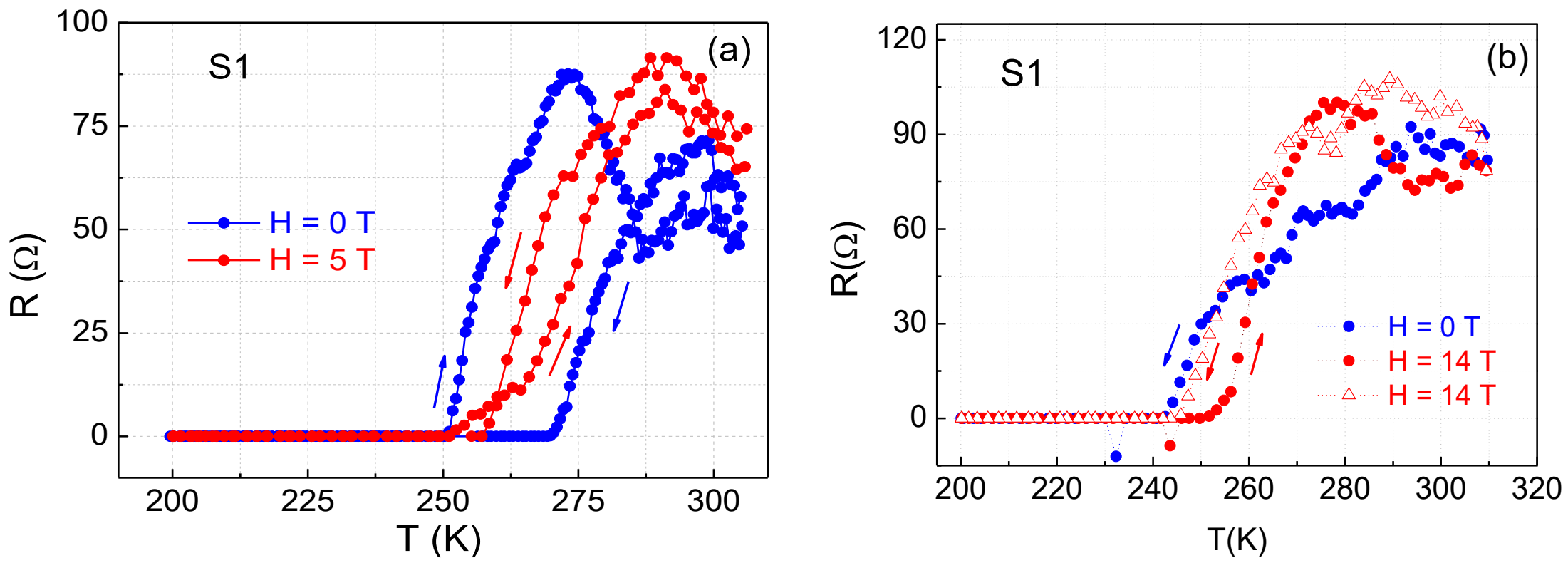

Fig 3: R versus T for Ag-Au nanostructured film on $SiO_2$/Si substrate in zero field as well as in magnetic field of H = 5 T (a) and H = 14 T (b). The results shows the large variation in the transition temperature in cooling and heating cycles, besides the inconsistence nature of transition in the applied magnetic field.

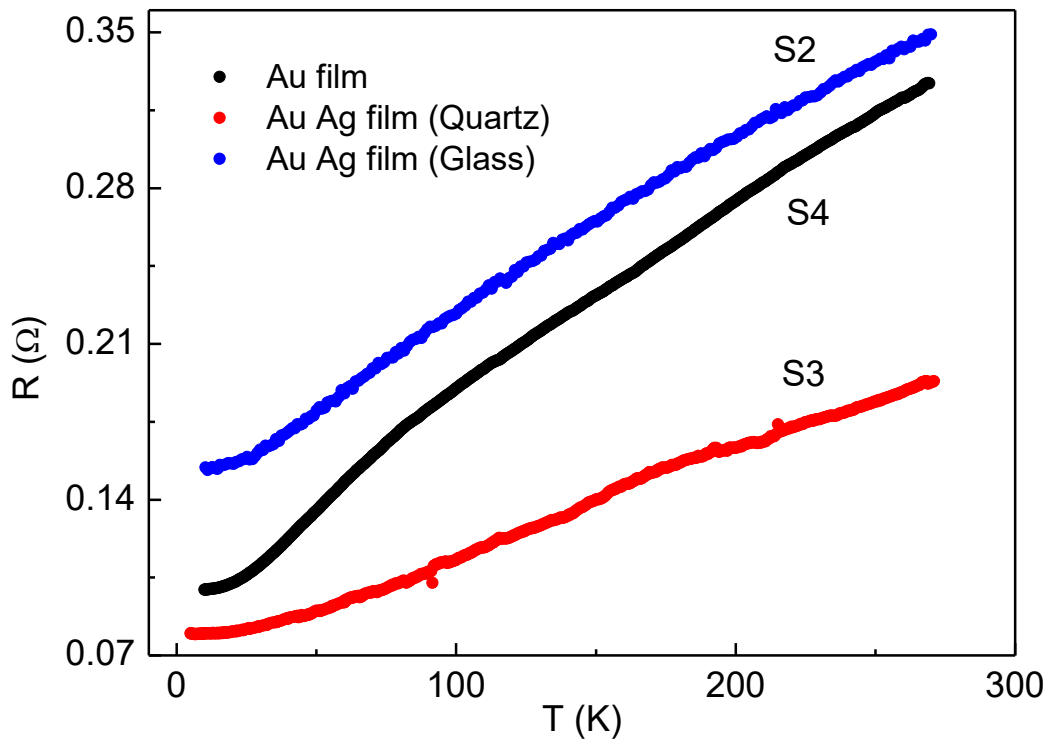

Fig 4: R versus T for Ag-Au nanostructured film on glass (S2) and quartz (S3) substrates and pure Au film on quartz substrate (S4), showing metallic behavior down to 10 K.

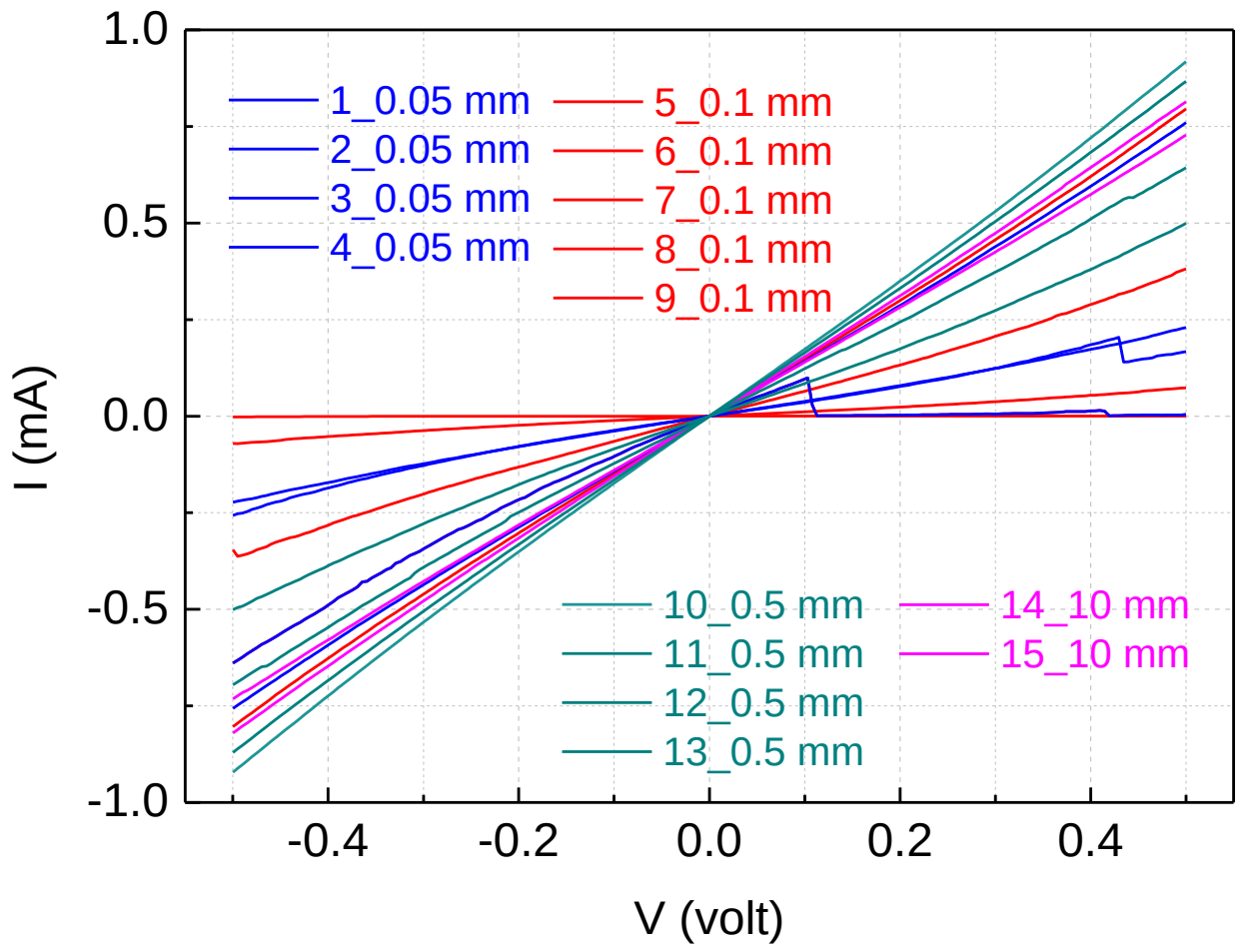

Fig. 5: Current- Voltage (I-V) characteristics of the film measured at room temperature with different electrodes (two probe) separations of 0.05 mm, 0.1 mm, 0.5 mm and 10 mm. Numbers 1 to 15 corresponds to the different regions of the nanostructured films.

Supplementary Information

# Observations of zero electrical resistance of Au-Ag thin films near room temperature

M. K. Hooda[1], P. Kumar[2], Viswanath Balakrishnan[2], and C. S. Yadav[1]

[1]School of Basic Sciences, Indian Institute of Technology Mandi, Mandi-175005 (H.P.) India

[2]School of Engineering, Indian Institute of Technology Mandi, Mandi-175005 (H.P.) India

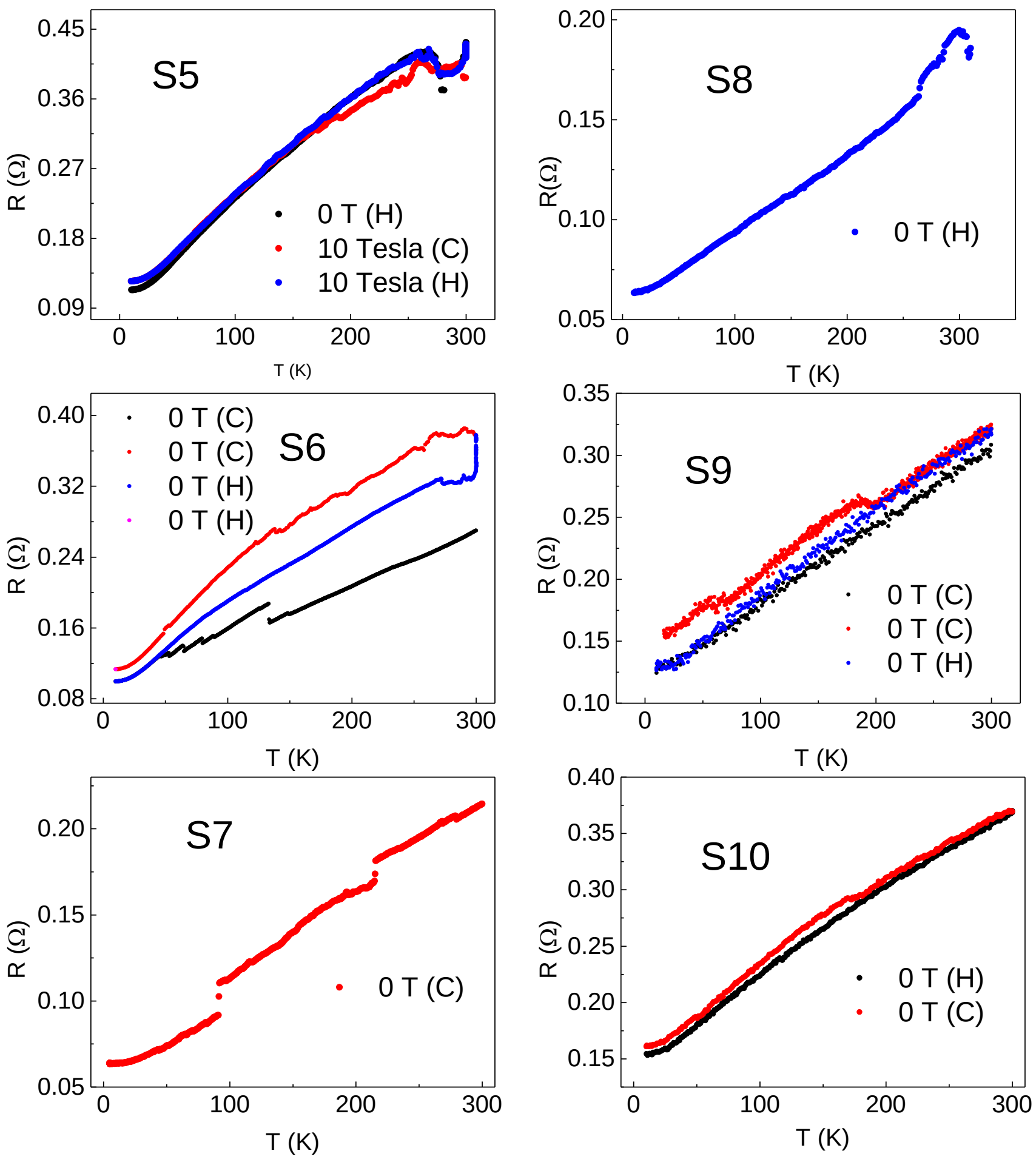

Fig. S1: The temperature dependence of electrical resistance of various nanostructured thin films (from S5 to S10) in heating (H) and cooling cycles (C). All the films show metallic behavior despite the abrupt changes in resistance.